# Continuous Time Random Walks and South Spain Seismic Series

Posadas, A. [1,2]; Morales, J.[2]; Vidal, F. [2, 3]; Sotolongo-Costa, O. [4];

Antoranz, J.C. [5]

[1] *Departamento de Física Aplicada. Universidad de Almería. Spain.*

[2] *Instituto Andaluz de Geofísica. Granada. Spain.*

[3] *Instituto Geográfico Nacional. Ministerio de Fomento. Madrid. Spain*

[4] *Departamento de Física Teórica. Universidad de la Habana. Cuba.*

[5] *Departamento de Física Matemática y Fluidos. UNED. Madrid. Spain*

## Abstract

Lévy flights were introduced through the mathematical research of the algebra or random variables with infinite moments. Mandelbrot recognized that the Lévy flight prescription had a deep connection to scale-invariant fractal random walk trajectories. The theory of Continuous Time Random Walks (CTRW) can be described in terms of Lévy distribution functions and it can be used to explain some earthquake characteristics like the distribution of waiting times and hypocenter locations in a seismic region. This paper checks the validity of this assumption analyzing three seismic series localized in South Spain. The three seismic series (Alborán, Antequera and Loja) show qualitatively the same behavior, although there are quantitative differences between them.

# INTRODUCTION

The current research about earthquakes exhibits some important concepts recently introduced in Physics: Self-organised criticality (SOC). This SOC is perhaps one of the concepts recently most used in this field. Fractal geometry, hierarchical models, depinning and Lévy distributions are probably less used though intimately connected with SOC.

Bak, Tang and Wiesenfeld (1987) proposed that spatio-temporal non-linear dynamical system, with quasi-static incoming and outcoming fluxes localized, for instance, at the borders, evolve spontaneously towards a stationary self-organised critical state. Neither length nor time scales others than those deduced from the size of the system and that of the elementary cell appear in the system. Subsequently it has been proposed that earthquakes may be an important natural phenomenon exhibiting SOC (Takayasu and Matsuzaki, 1988; Sornette and Sornette, 1989; Bak and Tang, 1989). These developments revived interest in much earlier work in which seismicity was modeled with blocks and springs. For instance, with a one dimension chain (Burridge and Knopoff, 1967), generating a series of studies involving numerical models of block-spring systems of various types (Carlson and Langer, 1989a, 1989b; Nakanishi, 1990; Brown, Scholz and Rundle, 1991). Scholz (1991) has argued that earth's entire crust is in a state of self-organised criticality. Thus the crust is everywhere on the brink of failure. Sornette (1991) has given similar arguments. Earthquakes may be the cleanest and most direct example of a self-organised critical phenomenon in Nature, and this is widely recognized

Another important characteristic of earthquakes, *i.e.*, the migration of hypocenters and its description in terms of anomalous diffusion has not received enough attention although it is closely related with criticality. This is the main goal of our paper.

The hypothesis of SOC for earthquakes leads to a power law for the temporal fluctuations for earthquake occurrence, which rationalize many observations. The Gutenberg-Richter law can be interpreted as a manifestation of the Self-organised critical behavior of the earth dynamics (Bak et al, 1988). Bak (1991) found that the exponents may differ from different models, but there is also the distinct possibility, known from equilibrium critical phenomena as *universality* that exponent depends only on geometrical and topological features such as the spatial dimension. Several groups have suggested that self-organised criticality is a natural explanation for the Gutenberg-Richter law (Sornette and Sornette, 1989; Bak and Tang, 1989; Ito and Matzusaki, 1990; Correig et al, 1997).

On the other hand, anomalous diffusion processes generally occur in disordered systems, *i.e.*, electronic conduction in amorphous semiconductors, atomic diffusion in glass like materials and others. The description of this phenomenon in terms of Lévy flights has shown to be adequate. (Vázquez, Sotolongo-Costa and Brouers, 1998). This kind of diffusion is usually modeled through a probability density of waiting times between successive steps in the walk, continuous time random walks (CTRW).

The theory of CTRW has been extensively developed (Bouchaud and Georges, 1990 and references therein). The existence of a wide distribution of waiting times leads to a subdiffusive regime where the mean square displacement grows slower than time. As we will show in this paper, some hypocenters are well described by a model of CTRW with a subdiffusive regime.

**DISTRIBUTION OF HYPOCENTERS AS A LÉVY FLIGHT**

It has been noted that the distribution of hypocenters exhibits fractal characteristics in its geographic representation in the seismic regions (Nakanishi *et al*, 1992). This kind of distribution could be modeled by some type of *anomalous* diffusion determined by some dynamics based on *waiting times*. Earthquakes can be considered as a relaxation mechanism of the earth crust loaded with inhomogeneous stresses, which accumulate at lithospheric-plate borders. This inhomogeneity determines an irregular distribution of hypocenters. Once a mainshock occurs, the landscape of the stresses on the earth crust redistributes itself to some extent, and a new event will occur when the accumulated stresses surpass again the threshold somewhere else (incidentally, these arguments remind the *punctuated equilibrium* behavior in evolution and many other natural phenomena). The new place of occurrence will be considered here as the new position of a random walk, which has to wait for a time $\tau_w$ on each site before the next jump. Once an earthquake has occurred somewhere, we can assume that the random walk has to wait until the redistribution of stresses leads to a new earthquake somewhere else, at a distance x from the place of the former event. This jump occurs suddenly, so that the waiting time $\tau_x$ for the walker to go from one point to another is much less than the waiting time $\tau_w$ in the place where the last earthquake has occurred. The waiting time is a random variable distributed according to a given law $p(\tau_w)$. We also assume that the waiting time is not correlated to the length of the jump x, distributed as p(x). The distributions $p(\tau_w)$ and p(x) for a given seismic region should differ. Indeed, since earthquakes occur mainly in some regions, we can suppose that the width of p(x) is small. This corresponds to values $\alpha_x = 2$, where $\alpha_x$ is the power of the tail of the distribution p(x), assuming that the tail is described by a power law ($p(x) \approx x^{a_x}$) *i.e.*, the distribution p(x) belongs to the attraction basin of a Gaussian law. This means that p(x) has finite variance. If we limit our analysis to a seismic region then we can imagine the hypocenter as a random walk confined in a given region with a wide distribution of waiting times. We adopt the CTRW model to describe the migration of earthquakes in a given seismic region. This standpoint is supported by the representation of earthquakes as the slipping between asperities, where displacements between blocks of a fault occur leading to the seism. In this representation, as we already pointed out, the epicenter is localized in the place where the two asperities collide. This model is a good tool to represent the migration of hypocenters in a seismic region as a problem of the diffusion of a random walk in a comb-like structure. So, CTRW analysis is directly applicable and has well known results for the distribution of waiting times and mean square displacement (Buochaud and Georges, 1990). The diffusion process is characterized by the scattering function $F(k,t)$, the Fourier transform of the diffusion front. Properties like the mean square displacement can be derived from this function, *i.e.*:

$$<x^2> = \partial^2 F / \partial k^2 \big|_{k=0} \tag{1}$$

Let $N$ be the number of steps performed by a walker during time $t$; $N$ is, in general, a random variable, which depends on the duration of the jumps and waiting times. The limit distribution for large $N$ will be a stable Lévy distribution (Feller, 1966; Gnedenko, 1954) and then we obtain (Sotolongo-Costa et al, 1999):

$$<x^2> \sim t^{a_w - a_x + 2} \tag{2}$$

As we have supposed that the geographic distribution of hypocenters belongs to the attraction basin of the Gauss law, this allows to put, in a rough approximation, $\alpha_x \cong 2$, obtaining:

$$<x^2> \sim t^{a_w} \tag{3}$$

This corresponds to a subdiffusive behavior since the wide character of waiting times implies small values for $\alpha_w$. To gain insight in the allowed values for $\alpha_w$ its is helpful to relate the nature of the rough profile of faults with the known problem of CTRW in comb-like structures. In that model the waiting time distribution function is (Bouchaud and Georges, 1990):

$$\Psi(t) \sim t^{(1+a_w)} \sim t^{1.5} \qquad (4)$$

If diffusion occurs in a comb-like structure as that evoked by the fault profile, this then implies $<x^2> \sim t^{0.5}$ for the mean square displacement.

To check the validity of these assumptions in the next section we will analyze the diffusion of hypocenters in several seismic series localized in the Southern Spain region.

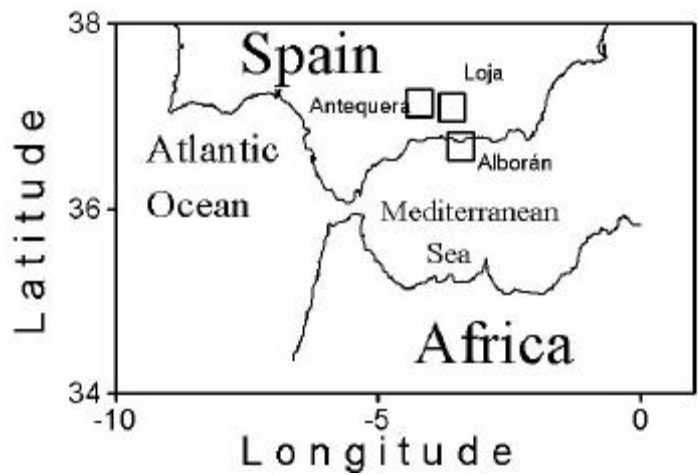

Figure 1. Situation map of the area under study. This region is located between Africa and Europe plates. Three square-boxes represent the epicentral area where Alboran Sea, Antequera and Loja Seismic Series occurred.

**SOUTHERN SPAIN SEISMIC ACTIVITY**

**Geological and Seismological Setting**

From the point of view of seismic activity, Southern Spain is the most dangerous region in Spain (Posadas, 1992). Located in the interaction zone between Europe and Africa, the area under study is situated in the central part of the Betic Cordilleras (fig.1). It contains the Granada Basin and several mountain ranges around it. Geologically,

Granada Basin is bounded to the north and to the west by Subbetic domain materials, mainly Jurassic and Cretaceous carbonate sedimentary series belonging to the Subiberic paleomarge. South and east sides are bounded by the Alpujarrides metamorphic units (schists, phyllites and quartzites of Palaeozoic and Triassic age and marbles of Triassic period) of the Alboran domain (García Dueñas and Balanyá, 1986; De Miguel et al, 1992). The structure of the crust is characterized by strong variations in the crustal thickness (Udías & Suriñach, 1980). The faulting present in the Granada Basin created a set of blocks that are structured at different levels that allow independent movements of them. There are also co-existing compressive and extensional deformations in nearby areas. These features fit into a general compressive framework, which produces contemporary extensional and compressive deformations on strike-slip faults. The seismogenetic areas are concentrated in three fracture systems having N10-30E, N30-60W and N70-100E directions. All the fracture systems are embedded in the Betic Area (Vidal, 1986; Peña *et al*, 1991; Posadas *et al*, 1993a; Posadas *et al*, 1993b).

**The Andalusian Seismic Network**

The Andalusian Seismic Network (located in Southern Spain) includes a microearthquake network, a network of accelerographs for strong and weak motions, an array device and a Broad-Band network. The microseismic network includes 18 stations, spread around the Andalusian territory, they are equipped with vertical sensors (Mark or Ranger) of 1 second natural period. The transmission of the ground motion is carried out in real time *via* radio to the Andalusian Institute of Geophysics located in the Cartuja Observatory (Granada, Spain) where an acquisition system under PC digitizes the recording to 100 samples per second with a dynamic range of 12 bits. An algorithm incorporated in the acquisition system, detects and records the earthquake on the hard disk. A monitorization of the seismicity of the Andalusian territory is done on analogical recording on printer paper.

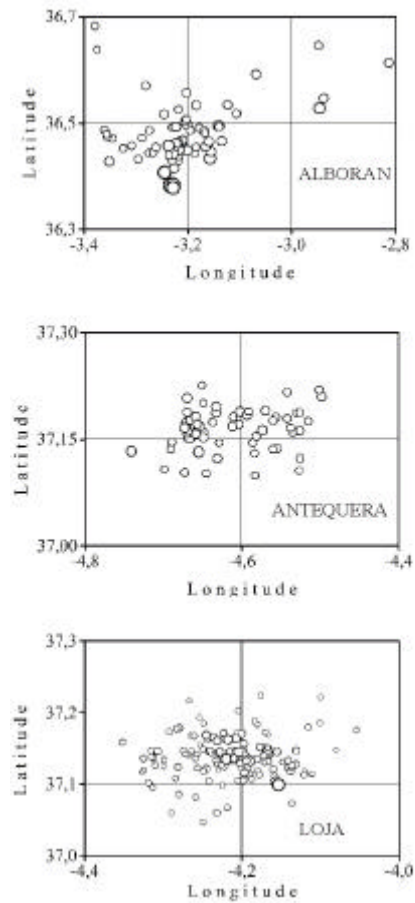

*Figure 2. The three square-boxes in figure 1 are showed in detail. In the upper figure it can be observed Alboran Sea Seismic Series; in the middle figure, Antequera Seismic Series is shown and, finally, Loja Seismic Series is presented in the lower figure. As greater circle radius are used as higher magnitude earthquake.*

**Seismic Activity**

The Andalusian Seismic Network belonging to the Andalusian Geophysics Institute provide us a wide seismic catalogue that involve more than fifteen thousands of earthquakes and microearthquakes. The majority of earthquakes have a low seismic magnitude (less than 5.0). However, catastrophic earthquakes occurred in the past with intensities (MSK scale) ranging from IX to X (Vidal, 1986). The Vera earthquakes (1406 and 1518) caused lots of damages in this town. The Almería 1522 earthquake destroyed buildings like the Alcazaba (Arab castle built about 955 *bc*. with more than 43000 m$^2$ of extension) and the cathedral. In 1804, the Dalias earthquake killed 300 people and all the buildings were completely destroyed (including the City Hall and the Church). The Andalusian 1884 Earthquake was the most important event in this area and serious damages were found in cities far away the epicenter zone.

The location feasibility of the network allow us to have high precision hypocenter determination (Serrano, 1999). The earthquake locations are most affected by errors in the determination of depth; this poor resolution is due to the trade-off between origin time and depth (Lomnitz, 1980) and also to the dependence of depth on the velocity model. The latter dependence is an important effect in Andalusian region (South Spain) where there are strong lateral gradients in crustal thickness (Hartzfeld & Boloix, 1978; Banda, 1988). For seismicity studies where the utmost precision possible in the determination of hypocenters is required, the use of an alternative method for testing the calculated depths is needed. In this paper we use the method derived by Wadati (Wadati, 1933) and applied to the Andalusian zone by Ibañez et al. (1987). Thus much more exact hypocentral localization was achieved. The Granada Basin has a high activity microearthquakes with hypocenters shallower than 20 km. Seismic activity in the area under study is characterized by the occurrence of seismic series and swarms. Three seismic series have been chosen to test if its spatial and temporal distributions are such that they can be described by a Lévy flight with anomalous diffusion (in this case in a subdiffusive regime).

*The Alboran Sea Seismic Series (1997-1998)*
The Alboran Sea Seismic Series occurred from June 1997 until January 1998. After the main shock (4.5 Richter magnitude), more than 300 microearthquakes took place between 2.8º and 3.4º W Longitude and 36.3º and 36.7º N Latitude (fig. 2). This zone is under the Mediterranean Sea in an area called Alboran Sea. Figure 3 (above) shows the Gutenberg-Richter relation for this seismic series; b parameter around -0.80 (±0.03, correlation factor ≈ 0.987) indicates us that probably our catalogue needs to incorporate magnitudes lower than 2.0, but actually this data are not yet available.

*The Antequera Seismic Series (1989)*

In the period June 7-9, 1989, 158 earthquakes of magnitude smaller than 3.4, occurred between 4.4º and 4.8º W Longitude and 36.9º and 37.3º N Latitude (fig. 2). This zone is near the village of Antequera (Málaga, Spain) and the population felt many of the earthquakes. Magnitude-frequency relation (fig. 3, middle) shows that the catalogue for the series is completed for earthquakes ranging from 1.5 to 3.4 magnitude (b = -1.15 ± 0.10 and correlation factor ≈ 0.934).

*The Loja Seismic Series (1985)*

A high activity of microearthquakes occurred during 1985 in several groups of faults N40-55W in the Granada Basin. Sierra Loja area was the most active zone in February; more than 300 microearthquakes and earthquakes took place in a small area between 4.1º and 4.3 º W longitude and between 37.1 º and 37.2 º N latitude (fig. 2). The main event had a magnitude 3.7 (Richter magnitude) and V degrees (MSK scale) of maximum intensity. The mechanisms calculated (Vidal, 1986) for the most relevant earthquakes show the above-mentioned alignment. For this study, about 125 well-located earthquakes have been used. The Gutenberg-Richter analysis (fig. 3, below) shows that the catalogue for the seismic series is complete from 1.5 to 3.7 magnitude (b = -1.07 ± 0.30 and correlation factor ≈ 0.992).

**RESULTS AND CONCLUSIONS**

Lévy flights were introduced through the mathematical research of the algebra or random variables with infinite moments. Mandelbrot recognized that the Lévy flight prescription had a deep connection to scale-invariant fractal random walk trajectories. We have considered the occurrence of earthquakes like jumps in space and in time; figure 3 (second and third columns) shows the result after to apply Lévy Flight Model to the three seismic series studied in this paper. Normalized distance *vs* time were used to depict units: *D* is the average distance between earthquakes and τ was chosen to be the average time between two consecutively earthquakes. It has been found (fig. 3, second column) that data can be modeled with a potential law as:

$$\frac{<d^2>}{<D^2>} \propto \left(\frac{t}{t}\right)^{a_w}$$

with $a_w \in [0.53, 0.66]$, indicating its subdiffusive behavior. The correlation factor vary from 0.90 to 0.98, indicating a good fit. These high correlation factors show that we have found a very good confirmation of equation (3) with $\alpha_w$ predicted by equation (4). Correspondingly, fig. 3 (third column) shows the waiting time distribution function, *i.e.*, the normalized time distribution between earthquakes for all the former processed events. The expected fit is in the form:

$$\log \frac{n}{N} = A + B \Delta t$$

with B around 1.5. The graphs show a logarithmic plot with a slopes ranging from 1.69 to 1.23; this fact corroborates in a satisfactory measure our standpoint.

The most important parameter in this paper is $a_w$. We have found its numerical value from an spatial-temporal analysis and also, from the time between events. Both ways fit

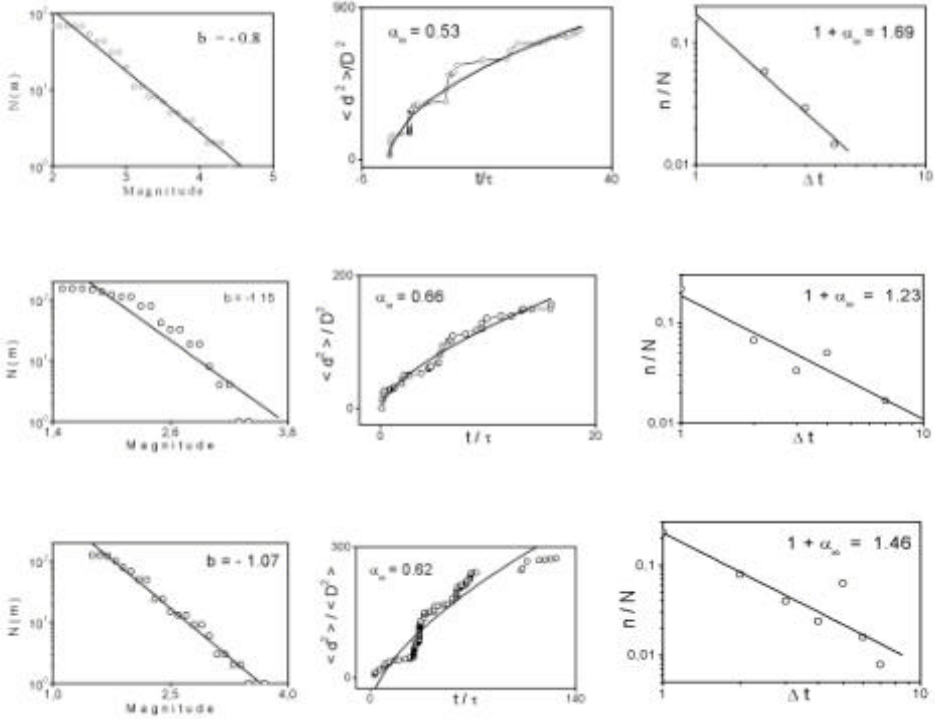

*Figure 3. Alboran Sea Seismic Series (above), Antequera Seismic Series (middle) and Loja Seismic Series (below) analysis. In all cases, Gutenberg-Richter analysis is in the first column; normalized distance vs time shows that $a_w$ agree with the theoretical prediction around 0.5 (second column); finally, the waiting time distribution function confirms our assumptions for $a_w$ value (third column).*

$a_w$ to the theoretical assumption. We have found an anomalous diffusive behavior (subdiffusive behavior) in the description of hypocenter migration from seismic series in South Spain. The three series present qualitatively the same behavior. This indicates that the system (the crust below three small areas in the South Spain Area) follows a non-equilibrium dynamics. The migration of earthquakes can be described as the diffusion of a walker in a comb-like structure, *i.e.*, it can be described with the CTRW model. These affirmations can be done for the three types of seismic series that can be found in South Spain and the subdiffusive behaviour is found in all the cases. Recently, Sotolongo-Costa et al (1999) have found a similar behaviour for the earthquakes in all of South Spain region using a catalogue with more than 7500 events. Moreover,

Vespignani et al (1995) have found a similar behaviour for the relaxation processes taking place after microfracturing of laboratory samples which give rise to ultrasonic acoustic emission signals; for instance, these researchers have got in their experiments that $1+a_w \approx 1.7$. So that, the scaling properties arise again: it doesn't matter if you are trying with a few earthquakes located in a small region or you are analyzing all the seismicity in a large zone.

The above mentioned results let us affirm that the concept of SOC provides a very stimulating framework within which to tackle the problem of defining and understanding the mechanics of the Lithosphere, *i.e.*, the connection between rupture at short time scale to that at large time scales and the geometric characteristics (scaling, Lévy distributions, etc.) essentially linked with it. These studies and others using probabilistic methods (Torcal et al, 1999; Torcal et al, 1999) allow us to a better understanding of how the crust in South Spain operates.

A Lévy flight description was used to characterize spatial and temporal distribution of earthquakes. Moreover, the exponent of the model for the mean square displacement has been quantitatively determined from observations and its value is around 0.5-0.6 with a high correlation factor. This value determines the temporal occurrence of earthquakes, as was here shown applying the results of CTRW model.

**Acknowledgements**

This work was partially supported by the CICYT project AMB99-1015-CO2-02, the DGESIC project HF1999-0129 (Almería University, Spain), the McyT project REN2001-2418-C04-02 and the Alma Mater prize given by the Havana University.